\documentclass[preprint,12pt]{elsarticle}

\usepackage{amssymb}

\journal{Materials......}

\begin{document}

\begin{frontmatter}



\title{Fabrication of self-powered photodetector materials based on Ni-doped ZnO/p-Si heterojunctions}


\author[inst1,inst2]{Eka Nurfani\corref{cor1}}
\ead{eka.nurfani@mt.itera.ac.id}  
\cortext[cor1]{Corresponding author}

\author[inst1]{Aldi Saputra}
\author[inst3]{Novalia Pertiwi}
\author[inst1,inst2]{Muhamad F. Arif}

\affiliation[inst1]{organization={Department of Materials Engineering, Faculty of Industrial Technology, Institut Teknologi Sumatera (ITERA)},
            city={South Lampung},
            postcode={35365}, 
            country={Indonesia}}

\affiliation[inst3]{organization={Center for Green and Sustainable Materials, Institut Teknologi Sumatera (ITERA)},
            city={South Lampung},
            postcode={35365}, 
            country={Indonesia}}

\affiliation[inst3]{organization={Department of Telecommunication Engineering, Faculty of Industrial Technology, Institut Teknologi Sumatera (ITERA)},
            city={South Lampung},
            postcode={35365}, 
            country={Indonesia}}

\begin{abstract}
In this paper, Ni-doped ZnO films were grown on a p-type silicon substrate via spray pyrolysis. The Ni dopant concentrations were varied by adjusting the weight ratio between Zinc Acetate Dehydrate (ZAD) and Nickel Chloride Hexahydrate (NCH), resulting in the ZnO, ZnO:Ni1$\%$, and ZnO:Ni3$\%$ samples. Field-effect scanning electron microscopy (FESEM) images revealed that Ni doping significantly reduced the nanostructure size from 326 nm (ZnO) to 146 nm (ZnO:Ni3$\%$). Similarly, X-ray diffraction (XRD) analysis also shows the decrease of the crystallite size with increasing Ni doping, from 44 nm (ZnO) to 35 nm (ZnO:Ni3$\%$). Current-voltage (I-V) measurements were conducted at a bias voltage of 0 and 5 V to examine electrical and self-powered photodetection properties. All samples demonstrate self-powered photodetector performance. At the bias of 0 V, the undoped ZnO exhibited a higher photo-to-dark-current ratio (162) as compared to ZnO:Ni1$\%$ (18) and ZnO:Ni3$\%$ (16). The ZnO:Ni3$\%$ samples displayed faster rise (0.4 s) and fall times (1.7 s) as compared to the pure ZnO (10.8 s for rise time and 9.1 s for fall time), highlighting their potential for applications requiring rapid photoresponse. The findings provide valuable insights into optimizing the performance of ZnO-based photodetectors through controlled Ni doping, enabling advancements in self-powered photodetection technology for energy-efficient optoelectronic devices.
\end{abstract}

%


\end{frontmatter}


\section{Introduction}
Photodetectors are vital in modern optoelectronic systems, allowing light conversion into electrical signals for applications ranging from environmental monitoring and biomedical imaging to optical communication and security systems \cite{Ma2024}. In recent years, the development of self-powered photodetectors has garnered significant attention because of their potential to operate without an external power source, thus improving energy efficiency and enabling integration into portable and autonomous systems. These devices are desirable for environments with limited energy, such as remote sensing and wearable electronics \cite{Zou2024}. Among the various materials explored for photodetectors, zinc oxide (ZnO) has emerged as a promising candidate because of its unique material properties. ZnO is a wide bandgap semiconductor (3.37 eV) with a high exciton binding energy of 60 meV, which ensures efficient light absorption and strong excitonic effects at room temperature \cite{KUSNAIDI2024, NURFANI2024}. ZnO is an ideal material for ultraviolet (UV) photodetectors owing to its chemical stability, low-cost fabrication, and compatibility with various substrates. However, pure ZnO-based photodetectors often face challenges such as slow response times and limited optical and electrical properties tunability, which restrict their application scope.

To address these limitations, doping ZnO with transition metals such as Mg \cite{NURFANI2024}, Mn \cite{KUSNAIDI2024}, Ti \cite{Nurfani2017_ZnOTi}, Cu \cite{NURFANI2021_ZnOCu}, Fe \cite{NURFANI2021_ZnOFe}, and Ni \cite{Bora2024} has been explored as an effective strategy. Doping is a well-established method for modifying the structural, electrical, and optical properties of ZnO, enabling its performance to be tailored for specific applications. Doping introduces localized states in the bandgap, modifies carrier dynamics, and alters the nanostructure morphology, leading to enhanced photodetector performance \cite{Boruah2018}. Ni-doped ZnO is particularly appealing because of its potential to achieve a balance between sensitivity and response speed, two critical parameters for high-performance photodetectors \cite{Elkamel2018}. However, the influence of Ni doping on self-powered photodetection properties remains underexplored, particularly in optimizing the trade-off between sensitivity and response speed for practical applications. 

In this study, using the spray pyrolysis method, we investigate the effect of Ni doping on the structural, electrical, and photodetection properties of ZnO films grown on p-type Si substrates. By varying the concentration of Ni dopant, we aim to elucidate the relationship between doping levels and key performance parameters, such as nanostructure size, crystallite size, sensitivity, and photoresponse speed. FESEM and XRD analyses characterize the structural changes induced by Ni doping, while I-V measurements under various bias voltages are used to evaluate the self-powered photodetection behavior. These findings provide valuable insights into designing and optimizing ZnO-based photodetectors, contributing to advancing energy-efficient optoelectronic devices.

\section{Experimental Methods}
\subsection{Synthesis of Ni-doped ZnO}
Silicon substrates (1 × 1 cm) were cleaned using a three-step ultrasonic process with acetone, ethanol, and distilled water for 10 minutes per solution. The substrates were then dried using a compressed air compressor to remove residual solvents. This cleaning process was performed to eliminate surface contaminants. Furthermore, Zinc acetate dihydrate (ZAD) and nickel(II) chloride hexahydrate (NCH) were used as precursors for Ni-doped ZnO. The precursor solutions were prepared by dissolving the appropriate amounts of ZAD and NCH in 40 mL of distilled water, namely ZnO, ZnO:Ni1$\%$, and ZnO:Ni3$\%$ samples. The solutions were stirred at 800 rpm using a magnetic stirrer on a hot plate at room temperature for 30 minutes to ensure homogeneity. The prepared solutions were deposited on the cleaned p-Si substrates using the spray pyrolysis method at a substrate temperature of 400$^\circ$C. the distance between the nozzle and substrate was kept at about 2 cm. The deposition time for each sample was 10 minutes. A schematic experiment was illustrated in \textbf{Figure \ref{fig:exp+XRD}(a)}.
\subsection{Characterizations}
Morphological analysis of the samples was carried out using a Thermo Scientific Quatro S SEM. This characterization was carried out at various magnifications of 10,000, 50,000, and 100,000 times. The crystallinity of the sample was carried out with XRD Panalytical Xpert 3 Powder using CuK$\alpha$ radiation with a wavelength of 1.5406 \r{A}. This characterization also ensures the fabricated sample was successfully deposited using the spray pyrolysis method by looking at the structural and morphological properties. The photodetector performance of the samples was characterized using a Source Measure Unit (SMU) Keithley 2450. To fabricate the device, silver electrodes ($4 \times 4$ mm) were deposited on the Ni-doped ZnO film and below the p-Si substrate to make a p-n junction. Current-voltage (I-V) and current-time (I-T) characterizations were carried out in the dark and under illumination with a 50 W halogen lamp.
\section{Results and Discussions}
\subsection{Crystal Structure}
\begin{figure}
    \centering
    \includegraphics[width=0.7\linewidth]{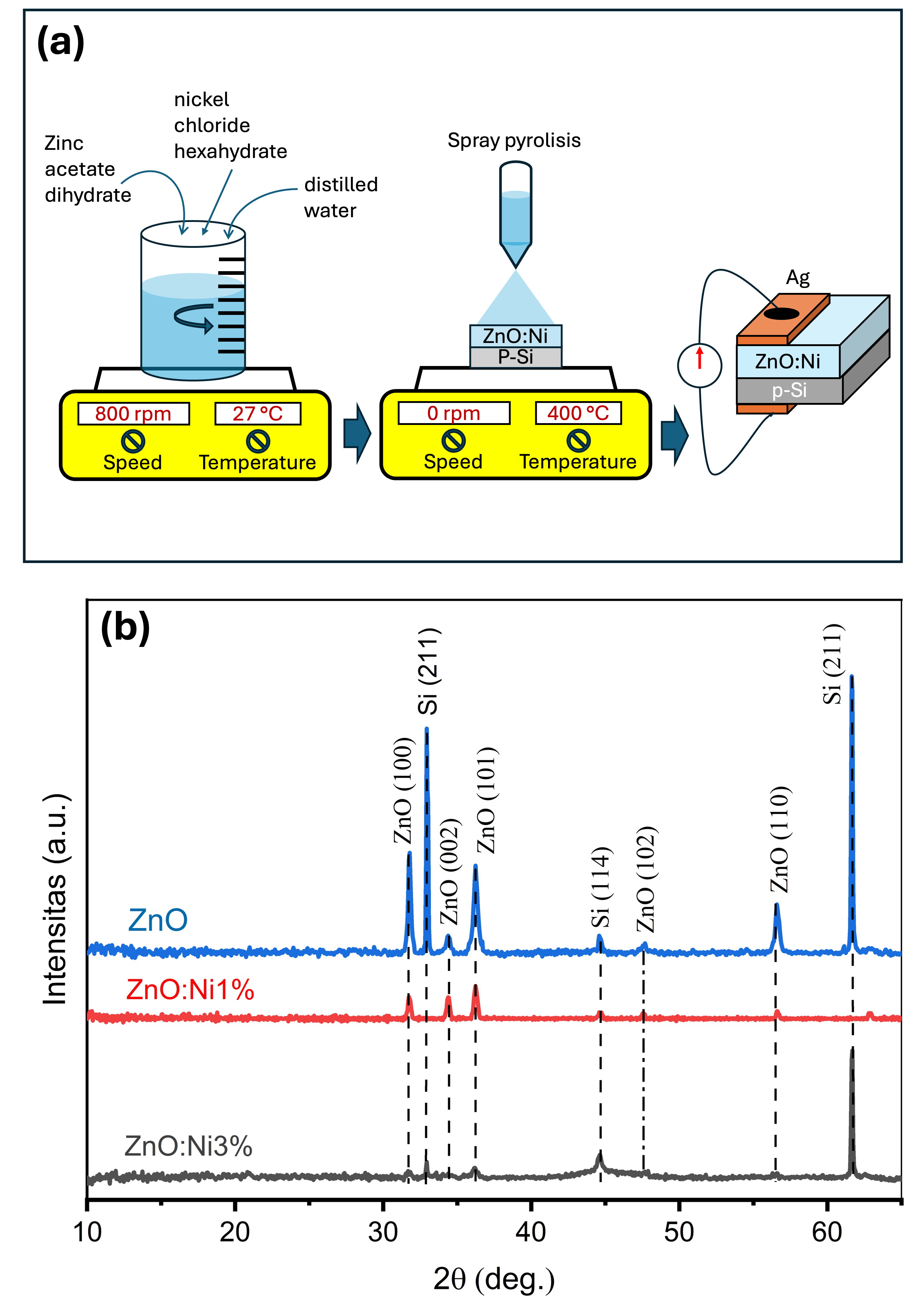}
    \caption{(a) Schematic diagram of the Ni-doped ZnO film deposited by spray pyrolisis and device fabrication for I-V curve measurements. (b) XRD pattern of Ni-doped ZnO films deposited on the p-Si substrate}
    \label{fig:exp+XRD}
\end{figure}

\textbf{Figure \ref{fig:exp+XRD}(b)} shows a diffractogram of the Ni-doped ZnO samples on the p-Si substrate. XRD analysis compared the diffraction peaks with the JCPDS 36-1451 (Powder Diffraction File) linked to the standard ZnO database. The resulting sample that has been fabricated has a hexagonal wurtzite crystal structure. No additional peaks correspond to Ni and NiO, indicating that the Ni doping is successfully inserted into the ZnO host lattice. The diffraction peaks are found at $2\theta$ of 31.77$^\circ$, 34.41$^\circ$, 36.24$^\circ$, 47.58$^\circ$, and 56.58$^\circ$, which correspond to the (100), (002), (101), (102), and (110) planes. Interplanar spacing ($d$) was calculated using the Bragg law
\begin{equation}
    d \sin{\theta} = n \lambda,
\end{equation}
The $d$ values for all samples are 2.81, 2.81, and 2.82 \r{A} for the ZnO, ZnO:Ni1$\%$, and ZnO:Ni3$\%$ samples, respectively. When Ni$^{2+}$ (ionic radii of 0.69 \r{A}) substitute Zn$^{2+}$ ions (ionic radii of 0.74 \r{A}) in the ZnO lattice, the smaller Ni ions introduce strain in the crystal lattice. This substitution distorts the lattice locally, slightly expanding along specific crystallographic directions, which may increase the interplanar spacing.

Furthermore, the crystallite size of the films ($L$) was calculated using the Scherrer equation 
\begin{equation}
    L =  \frac{K \lambda}{\beta \sin{\theta}}
\end{equation}
where $K$ is the Scherrer constant, $\lambda$ is wavelength of X-ray, and $\beta$ is the full-width at half maximum. The crystallite size calculated from the Scherrer equation was 44, 37, and 35 nm for the ZnO, ZnO:Ni1$\%$, and ZnO:Ni3$\%$ samples, respectively. 
\subsection{Morphological Analysis}
\begin{figure}
    \centering
    \includegraphics[width=\linewidth]{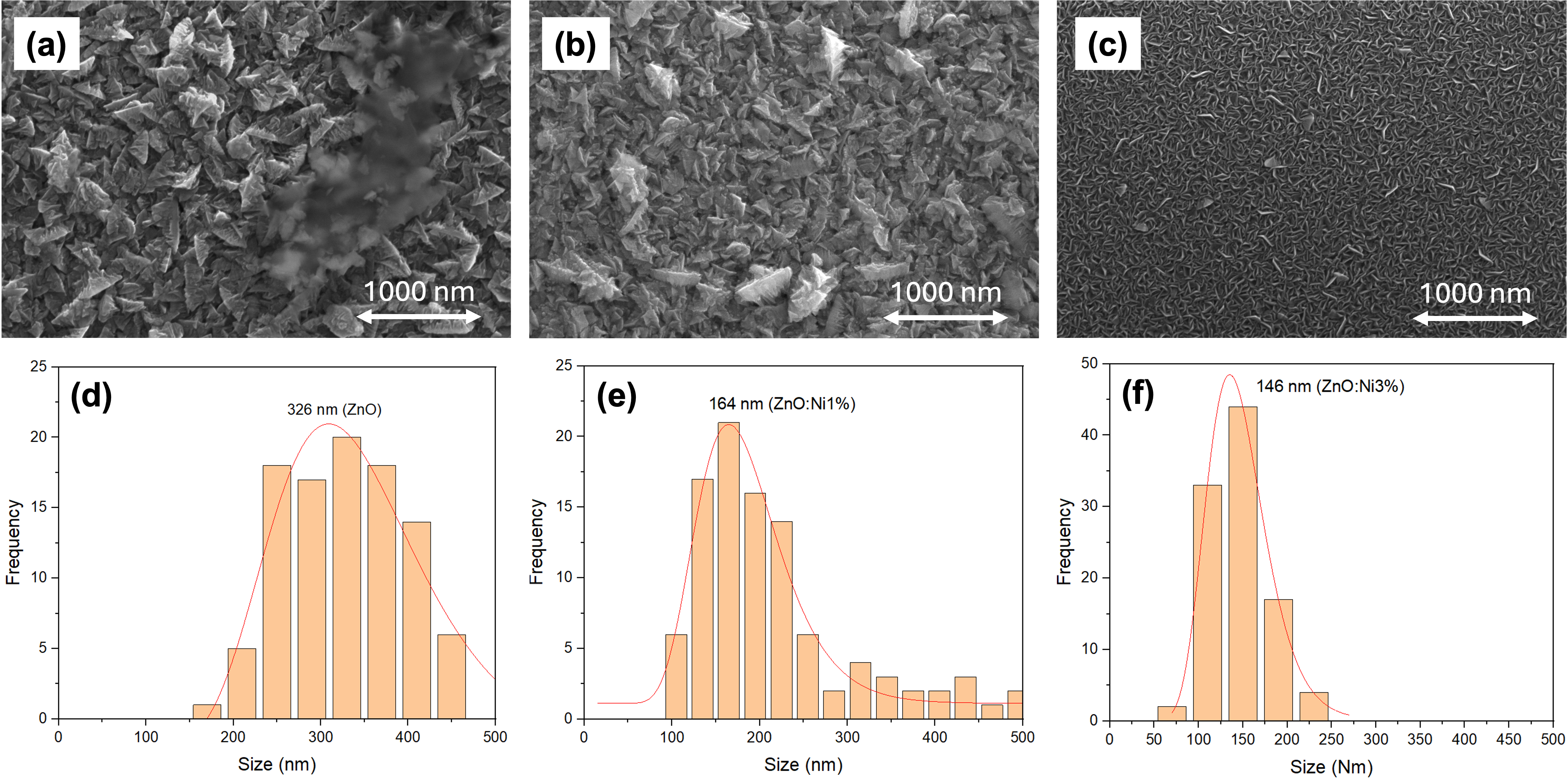}
    \caption{(a-c) Surface morphological images and (d-f) normal distribution of particle size for the ZnO, ZnO:Ni1$\%$, ZnO:Ni3$\%$ samples grown on the p-Si substrate.}
    \label{fig:Fig-SEM}
\end{figure}

\textbf{Figure \ref{fig:Fig-SEM}} presents the SEM characterization results, showing a surface morphological image at 100,000x magnification and the average size distribution of the nanostructured Ni-doped ZnO fabricated using the spray pyrolysis technique. The morphology of pure ZnO resembles flower petal fragments with overlapping flake layers, as also reported by another group \cite{Fattah-2024}. The observed structures appear larger and fragmented.  After Ni doping, the structures become finer and smaller than pure ZnO, possibly indicating growth inhibition due to the addition of Ni \cite{Shirage-2016}. By using 100 samplings, the estimated size of the nanostructure is 326 nm (ZnO), 164 nm (ZnO:Ni1$\%$), and 146 nm (ZnO:Ni3$\%$). This is consistent with the effect of metal doping such as Ni which often inhibits crystal growth, resulting in smaller structures. Ni$^{2+}$ ions (radius of 0.69 \r{A}) can replace Zn$^{2+}$ ions (radius of 0.74 \r{A}) in the ZnO crystal lattice. The presence of Ni$^{2+}$ ions on the particle surface or in the reaction solution can interfere with the nucleation of ZnO. This causes a decrease in the crystal growth rate because Ni acts as a blocker at the active growth sites.

\subsection{Performance of Photodetector}
\begin{figure}
    \centering
    \includegraphics[width=\linewidth]{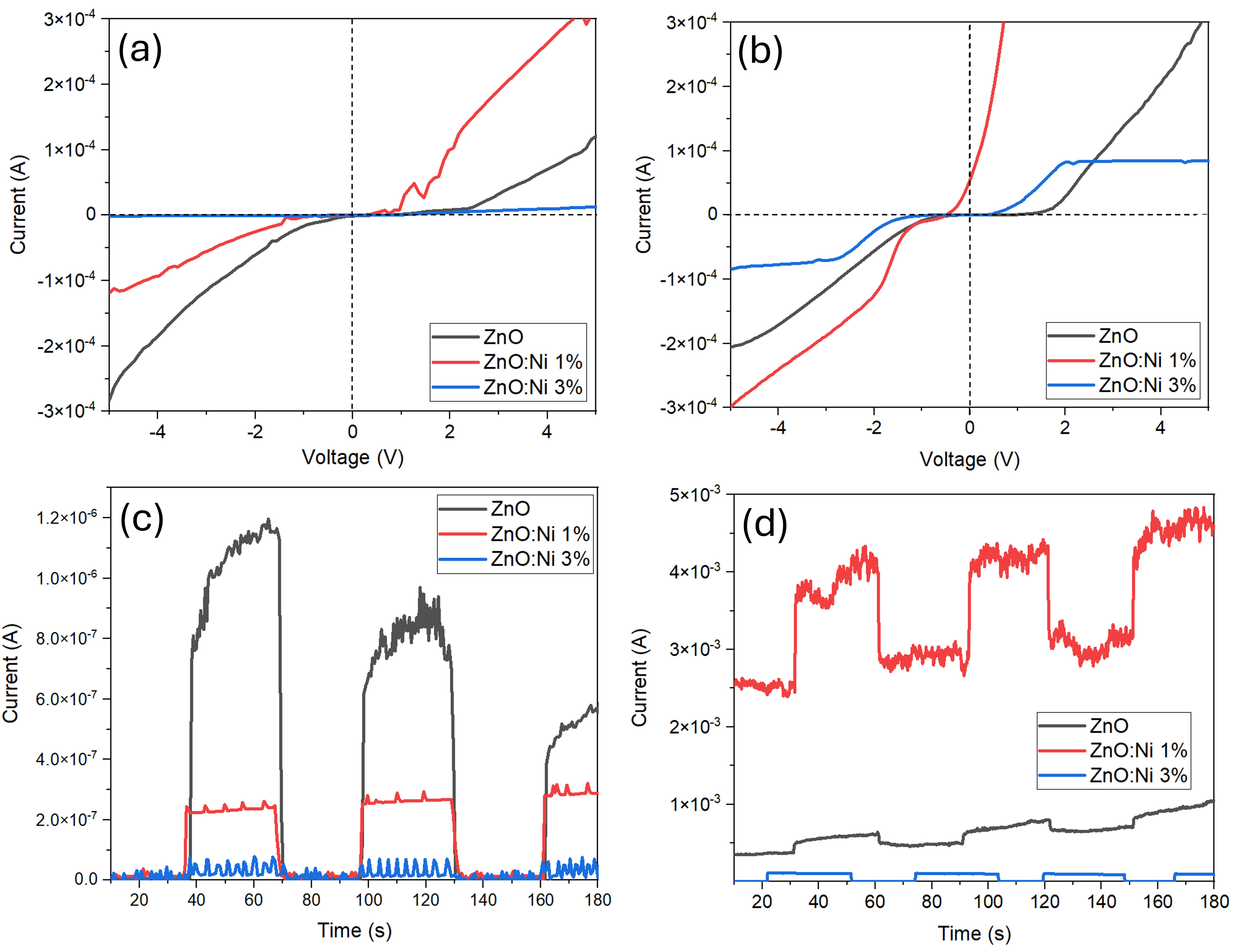}
    \caption{I-V curves of all samples under the (a) dark and (b) illumination conditions. I-T curves under the dark and illumination for all samples under the bias voltage of (c) 0 V and (d) 5 V.}
    \label{fig:IV}
\end{figure}

\textbf{Figure \ref{fig:IV}(a-b)} is I-V curves under the dark and illumination conditions. For all samples, the photocurrent ($I_p$) is higher than the dark current ($I_d$), indicating a photodetector behavior. Lower dark current is caused by oxygen adsorption by free electrons generated by oxygen vacancies at the surface of ZnO film ($O_2 + e^- \rightarrow O_2^-$) \cite{Girard_2018}. Please note that the oxygen vacancies in the ZnO system are the origin of n-type conductivity \cite{Liu-2016}. This adsorption induces the formation of a depletion layer that blocks the current flow \cite{Boruah-2019}, resulting in a low current. ZnO with a smaller nanostructure size has higher oxygen vacancies on its surface, resulting in a larger depletion layer and low dark current as shown in the ZnO:Ni3$\%$. By illuminating the sample with photon energy higher than the ZnO bandgap, electrons from the O-2p valence band are excited to the Zn-3d conduction band, leaving holes in the valence band. The holes move to the surface to discharge the negatively charged adsorbed oxygen ion ($O_2^- + h^+ \rightarrow O_2$), releasing the $O_2$ molecules \cite{Girard_2018}. The release of adsorbed oxygen molecules increases the conductivity during the illumination condition.

The dark current ($I_d$) and photocurrent ($I_p$) of the ZnO:Ni$1\%$ sample show the highest values compared to the ZnO and ZnO:Ni$3\%$ samples. Interestingly, a high photocurrent value in the doped sample at a bias voltage of 0 V indicates photovoltaic characteristics. Moreover, I-T curves were measured to investigate the photovoltaic behavior and response time at the bias voltage of 0 and 5 V, as shown in \textbf{Figure \ref{fig:IV}(c-d)}. At the bias voltage of 0 V, the $I_d$ values are $7.4 \times 10^{-9}$ A, $1.4 \times 10^{-8}$ A, and $4.8 \times 10^{-9}$ A, while the $I_p$ values are $1.2 \times 10^{-6}$ A, $2.5 \times 10^{-7}$, and $7.6 \times 10^{-8}$, for ZnO, ZnO:Ni$1\%$, and ZnO:Ni$3\%$ samples, respectively. At the bias voltage of 5 V, the observed $I_d$ values are $3.7 \times 10^{-4}$ A (ZnO), $2.5 \times 10^{-3}$ A (ZnO:Ni$1\%$), and $1.9 \times 10^{-7}$ A (ZnO:Ni$3\%$). The observed $I_p$ values are $6.1 \times 10^{-4}$ A (ZnO), $4.1 \times 10^{-3}$ A (ZnO:Ni$1\%$), and $1.1 \times 10^{-4}$ A (ZnO:Ni$3\%$). Furthermore, the photo-to-dark current ratio or PDCR (${I_p}/{I_d}$) is a measure of the performance of the photodetector that is related to its sensitivity. A high PDCR means the photodetector can distinguish faint optical signals from background noise. The calculated PDCRs are 1.6, 1.6, and 578.9 at the bias voltage of 5 V, and 162.2, 17.9, and 15.8 at the bias voltage of 0 V for the ZnO, ZnO:Ni$1\%$, and ZnO:Ni$3\%$ samples, respectively. 

Interestingly, the observed higher PDCR in the pure ZnO at the bias voltage of 0 V and in ZnO:Ni$3\%$ at the bias voltage of 5 V can be attributed to several factors. At the bias of 0 V, the ZnO:Ni$3\%$ photodetector operates under equilibrium conditions, where the dark current is minimal. The absence of an external electric field results in a lower photocurrent, leading to a smaller PDCR. Applying the bias of 5 V to Ni-doped ZnO enhances the internal electric field, accelerating the separation and transport of photogenerated carriers. This results in a higher photocurrent and, consequently, a higher PDCR. So, the higher PDCR in pure ZnO at 0 V bias and in Ni-doped ZnO at 5 V bias is influenced by the interplay between bias voltage, doping effects, and structural modifications, which collectively impact the photodetector's performance. Furthermore, the responsivity ($R$) and detectivity ($D$) of the photodetectors can be calculated using the following equation,
\begin{equation}
    R = \frac{I_p - I_d}{PA}
\end{equation}
\begin{equation}
    D = R \sqrt{\frac{A}{2e I_d}}
\end{equation}
where $P$ is the power intensity, $A$ is the active area, and $e$ is the electronic charges. The calculated $R$ values at the bias voltage of 0 V are 6 mA/W (ZnO), 1 mA/W (ZnO:Ni1$\%$), and 0.4 mA/W (ZnO:Ni1$\%$). On the other hand, the calculated $D$ values are $2 \times 10^8$ (ZnO), $4 \times 10^7$ (ZnO:Ni1$\%$), and $2 \times 10^7$ (ZnO:Ni1$\%$) Jones. Thus, Ni doping provokes the decrease of the PDCR, responsivity, and detectivity of the detector at the bias of 0 V\\

Moreover, rise ($\tau_r$) and fall ($\tau_f$) times are critical parameters in evaluating the temporal response and performance of photodetectors through I-T curve analysis in \textbf{Figure \ref{fig:IV}(c-d)}. These parameters reflect the ability of the device to accurately and rapidly respond to changes in the incident optical signal, making them essential for applications requiring high-speed detection, such as optical communication, imaging, and sensing systems. The rise time is the time needed for the output of a photodetector to increase from 10$\%$ to 90$\%$ of its final steady-state value in response to an instantaneous optical step input. 
\begin{equation}
    I(t) = I_0 \left( 1 - \exp{ \left( \frac{-t}{\tau_r} \right) } \right)
\end{equation}
The fall time is the time required for the output of a photodetector to decrease from 90$\%$ to 10$\%$ of its maximum steady-state value after the optical input is abruptly turned off. 
\begin{equation}
    I(t) = I_0  \exp{\left( \frac{-t}{\tau_d} \right)} 
\end{equation}
At the bias voltage of 0 V, the $\tau_r$ values are 10.8, 0.5, and 0.5 s, while the $\tau_f$ values are 9.1, 1.8, and 1.8 s for the ZnO, ZnO:Ni$1\%$, and ZnO:Ni$3\%$ samples, respectively. By using a bias voltage of 5 V, the observed $\tau_r$ is 16.1, 0.2, and 0.1 s, while the $\tau_f$ value is 4, 0.3, and 0.5 s for the ZnO, ZnO:Ni$1\%$, and ZnO:Ni$3\%$ samples, respectively. Faster rise and fall times are observed in the Ni-doped ZnO under different bias voltages of 0 V and 5 V. Doping ZnO with Ni atoms introduces additional free carriers, which can enhance the conductivity and carrier mobility of ZnO \cite{Elkamel2018}. This improvement facilitates faster rise and fall times in photodetectors. Doping ZnO can introduce energy levels within its bandgap, which act as recombination centers for the photogenerated carriers. While too many recombination centers can negatively affect the performance of the photodetector, an optimal concentration helps facilitate the rapid recombination of carriers when the light source is turned off, resulting in a shorter fall time \cite{Lavanya-2024}. The photodetector parameters, compared with other published works, are summarized in \textbf{Table \ref{tab:PD-perform} and \ref{tab:compare}}. 

\begin{table}
    \centering
    \begin{tabular}{|c|c|c|c|c|c|c|}
\hline
 Samples   & $I_p / I_d$ [0V]  & $\tau_r$ (s) [0V] & $\tau_f$ (s) [0V] & $I_p / I_d$ [5V]  & $\tau_r$ (s) [5V]  & $\tau_f$ (s) [5V] \\
\hline         
     ZnO            & 162   & 10.8          & 9.1           &  1.6  & 16.1          & 4  \\
     ZnO:Ni1$\%$    & 18    & 0.5           & 1.8           &  1.6  & 0.2           & 0.3   \\
     ZnO:Ni3$\%$    & 16    & 0.4           & 1.7           &  579  & 0.1           & 0.2   \\
\hline
    \end{tabular}
    \caption{Photodetector parameters, such as PDCR ($I_p / I_d$), rise time ($\tau_r$), and fall time ($\tau_f$), under bias voltage of 0 and 5 V.}
    \label{tab:PD-perform}
\end{table}

\begin{table}
    \centering
    \begin{tabular}{|c|c|c|c|c|c|c|}
    \hline
    Materials  & Light & $R$ (A/W) & $D$ (Jones) & $\tau_r$ (s) & $\tau_f$ (s) & Ref. \\
    \hline
    Au/ZnO & 325 nm &  $5 \times 10^ {-4} $ & $27 \times 10^{10} $ & 0.012 & 0.012 &  \cite{Meng-2022}  \\
    CuO/TiO$_2$/p-Si & 300 nm & $2 \times 10^{-11}$ & $2 \times 10^{6}$ & 0.086 & 0.078 & \cite{GHOSH-2024} \\
    ZnO/GaN& 350 nm & $225 \times 10^{-3}$ & $5 \times 10^{13}$ & 0.976 & 0.042  & \cite{Mishra-2019} \\
    ZnO/Co$_3$O$_4$& UV-Vis & 0.021 & $4 \times 10^{12}$ & 6 & - & \cite{Ghamgosar-2019}\\
    ZnO/CuI &  365 nm & 0.061 & $2 \times 10^{10} $ & 0.41 & 0.08 & \cite{CAO-2021} \\
    ZnO/NiO & 355 nm & $4 \times 10^{-4}$ & - & 0.23 & 0.21 & \cite{Shen-2015} \\
    ZnO/Cu$_2$O & 370 nm & 0.24 & - & 0.02 & 0.03 & \cite{Lin-2020} \\
    ZnO/ZnS & 365 nm & 0.03 & $9 \times 10^{12} $  & 22.5 & 45 & \cite{Benyahia-2021} \\
    ZnO:Mn/p-Si& halogen & - & - & 0.4 & 1.2 & \cite{KUSNAIDI2024} \\
    ZnO:Ni/p-Si& halogen& $10^{-3}$ & $4 \times 10^7$ & 0.4 & 1.7 & This work \\

    \hline
    \end{tabular}
    \caption{Comparison of the self-powered photodetector performance at the bias voltage of 0 V with other published works}
    \label{tab:compare}
\end{table}

\section{Conclusions}
This study demonstrates the significant impact of Ni doping on the structural, electrical, and photodetection properties of ZnO films grown via the spray pyrolysis method. Ni doping effectively reduces the nanostructure and the crystallite sizes, as evidenced by FESEM and XRD analyses. The photodetection performance of the ZnO-based devices was evaluated under self-powered conditions, revealing that undoped ZnO exhibits higher sensitivity. In contrast, Ni-doped ZnO samples demonstrate faster rise and decay times, making them suitable for applications requiring rapid photoresponse. These findings provide valuable insights into optimizing ZnO-based photodetectors through controlled doping strategies, paving the way for developing energy-efficient optoelectronic devices with tailored performance characteristics.


\section*{Acknowledgement}
EN thanks the Ministry of Higher Education, Science, and Technology (Kemendiktisaintek), the Indonesia Endowment Fund for Education (LPDP), and Sumatera Institut (ITERA) for their research funding.

\section*{Conflict of Interest Statement}
The authors have no conflicts of interest to declare.

\section*{CRediT Author Statement}
\textbf{Eka Nurfani:} Conceptualization, Methodology, Resources, Writing - Original Draft, Supervision, and Funding Acquisition.
\textbf{Aldi Saputra:} Formal analysis, Investigation, and Writing - Original Draft.
\textbf{Novalia Pertiwi:} Resources and Writing - Review and Editing
\textbf{Muhamad F. Arif:} Writing - Review and Editing, and Supervision.

 \bibliographystyle{elsarticle-num} 
 \bibliography{cas-refs}

\begin{thebibliography}{10}
\expandafter\ifx\csname url\endcsname\relax
  \def\url#1{\texttt{#1}}\fi
\expandafter\ifx\csname urlprefix\endcsname\relax\def\urlprefix{URL }\fi
\expandafter\ifx\csname href\endcsname\relax
  \def\href#1#2{#2} \def\path#1{#1}\fi

\bibitem{Ma2024}
T.~Ma, N.~Xue, A.~Muhammad, G.~Fang, J.~Yan, R.~Chen, J.~Sun, X.~Sun, Recent
  progress in photodetectors: From materials to structures and applications,
  Micromachines 15~(10) (2024).
\newblock \href {https://doi.org/https://doi.org/10.3390/mi15101249}
  {\path{doi:https://doi.org/10.3390/mi15101249}}.

\bibitem{Zou2024}
J.~Zou, S.~Zhang, X.~Tang, Recent advances in organic photodetectors, Photonics
  11~(11) (2024).
\newblock \href {https://doi.org/10.3390/photonics11111014}
  {\path{doi:10.3390/photonics11111014}}.

\bibitem{KUSNAIDI2024}
R.~Kusnaidi, W.~S. Sipahutar, N.~Pertiwi, R.~Marlina, E.~Nurfani, Enhancement
  in zno-based self-powered photodetector by inserting mn dopant, Physica B:
  Condensed Matter 695 (2024) 416543.
\newblock \href {https://doi.org/https://doi.org/10.1016/j.physb.2024.416543}
  {\path{doi:https://doi.org/10.1016/j.physb.2024.416543}}.

\bibitem{NURFANI2024}
E.~Nurfani, L.~Nulhakim, D.~M. Muhammad, M.~Rozana, W.~Astuti, The enhanced
  sensing performance of zno-based photodetector by mg doping, Optical
  Materials 148 (2024) 114948.
\newblock \href {https://doi.org/https://doi.org/10.1016/j.optmat.2024.114948}
  {\path{doi:https://doi.org/10.1016/j.optmat.2024.114948}}.

\bibitem{Nurfani2017_ZnOTi}
E.~Nurfani, N.~Zuhairah, R.~Kurniawan, S.~Muhammady, I.~M. Sutjahja, T.~Winata,
  Y.~Darma, Influence of ti doping on the performance of a zno-based
  photodetector, Materials Research Express 4~(2) (2017) 024001.
\newblock \href {https://doi.org/10.1088/2053-1591/aa5773}
  {\path{doi:10.1088/2053-1591/aa5773}}.

\bibitem{NURFANI2021_ZnOCu}
E.~Nurfani, W.~Kesuma, A.~Lailani, M.~Anrokhi, G.~Kadja, M.~Rozana,
  W.~Sipahutar, M.~Arif, Enhanced uv sensing of zno films by cu doping, Optical
  Materials 114 (2021) 110973.
\newblock \href {https://doi.org/https://doi.org/10.1016/j.optmat.2021.110973}
  {\path{doi:https://doi.org/10.1016/j.optmat.2021.110973}}.

\bibitem{NURFANI2021_ZnOFe}
E.~Nurfani, A.~Lailani, W.~Kesuma, M.~Anrokhi, G.~Kadja, M.~Rozana, Uv
  sensitivity enhancement in fe-doped zno films grown by ultrafast spray
  pyrolysis, Optical Materials 112 (2021) 110768.
\newblock \href {https://doi.org/https://doi.org/10.1016/j.optmat.2020.110768}
  {\path{doi:https://doi.org/10.1016/j.optmat.2020.110768}}.

\bibitem{Bora2024}
A.~Bora, J.~George, Y.~Sivalingam, S.~Velappa~Jayaraman, G.~Magna, M.~S. R.~N.
  Kiran, L.~Vesce, R.~Paolesse, C.~Di~Natale, Self-powered photodetectors with
  nickel-doped zno nanorods for operation in low-light environments, ACS
  Applied Nano Materials 7~(8) (2024) 9324--9336.
\newblock \href {https://doi.org/10.1021/acsanm.4c00725}
  {\path{doi:10.1021/acsanm.4c00725}}.

\bibitem{Boruah2018}
B.~Deka~Boruah, S.~Naidu~Majji, S.~Nandi, A.~Misra, Doping controlled
  pyro-phototronic effect in self-powered zinc oxide photodetector for
  enhancement of photoresponse, Nanoscale 10 (2018) 3451--3459.
\newblock \href {https://doi.org/10.1039/C7NR08125A}
  {\path{doi:10.1039/C7NR08125A}}.

\bibitem{Elkamel2018}
I.~Ben~Elkamel, N.~Hamdaoui, A.~Mezni, R.~Ajjel, L.~Beji, High responsivity and
  1/f noise of an ultraviolet photodetector based on ni doped zno
  nanoparticles, RSC Adv. 8 (2018) 32333--32343.
\newblock \href {https://doi.org/10.1039/C8RA05567J}
  {\path{doi:10.1039/C8RA05567J}}.

\bibitem{Fattah-2024}
E.~M. Abdel-Fattah, S.~M. Alshehri, S.~Alotibi, M.~Alyami, D.~Abdelhameed,
  Hydrothermal synthesis of zno nanoflowers: Exploring the relationship between
  morphology, defects, and photocatalytic activity, Crystals 14~(10) (2024).
\newblock \href {https://doi.org/10.3390/cryst14100892}
  {\path{doi:10.3390/cryst14100892}}.

\bibitem{Shirage-2016}
P.~M. Shirage, A.~K. Rana, Y.~Kumar, S.~Sen, S.~G. Leonardi, G.~Neri, Sr- and
  ni-doping in zno nanorods synthesized by a simple wet chemical method as
  excellent materials for co and co2 gas sensing, RSC Adv. 6 (2016)
  82733--82742.
\newblock \href {https://doi.org/10.1039/C6RA15891A}
  {\path{doi:10.1039/C6RA15891A}}.

\bibitem{Girard_2018}
J.-P. Girard, L.~Giraudet, S.~Kostcheev, B.~Bercu, T.~J. Puchtler, R.~A.
  Taylor, C.~Couteau, Mitigating the photocurrent persistence of single zno
  nanowires for low noise photodetection applications, Nanotechnology 29~(50)
  (2018) 505207.
\newblock \href {https://doi.org/10.1088/1361-6528/aae417}
  {\path{doi:10.1088/1361-6528/aae417}}.

\bibitem{Liu-2016}
L.~Liu, Z.~Mei, A.~Tang, A.~Azarov, A.~Kuznetsov, Q.-K. Xue, X.~Du, Oxygen
  vacancies: The origin of $n$-type conductivity in zno, Phys. Rev. B 93 (2016)
  235305.
\newblock \href {https://doi.org/10.1103/PhysRevB.93.235305}
  {\path{doi:10.1103/PhysRevB.93.235305}}.

\bibitem{Boruah-2019}
B.~Deka~Boruah, Zinc oxide ultraviolet photodetectors: rapid progress from
  conventional to self-powered photodetectors, Nanoscale Adv. 1 (2019)
  2059--2085.
\newblock \href {https://doi.org/10.1039/C9NA00130A}
  {\path{doi:10.1039/C9NA00130A}}.

\bibitem{Lavanya-2024}
S.~Lavanya, T.~{Rajesh Kumar}, B.~Prakash, R.~{Rimal Isaac}, I.~Ashraf,
  Siddhartha, M.~Shkir, L.~Kansal, H.~Payal, S.~S. Sehgal, Effect of bi doping
  on the opto-electronic properties of zno nanoparticles for photodetector
  applications, Journal of Photochemistry and Photobiology A: Chemistry 446
  (2024) 115119.
\newblock \href
  {https://doi.org/https://doi.org/10.1016/j.jphotochem.2023.115119}
  {\path{doi:https://doi.org/10.1016/j.jphotochem.2023.115119}}.

\bibitem{Meng-2022}
J.~Meng, Q.~Li, J.~Huang, C.~Pan, Z.~Li, Self-powered photodetector for
  ultralow power density uv sensing, Nano Today 43 (2022) 101399.
\newblock \href {https://doi.org/https://doi.org/10.1016/j.nantod.2022.101399}
  {\path{doi:https://doi.org/10.1016/j.nantod.2022.101399}}.

\bibitem{GHOSH-2024}
C.~Ghosh, A.~Dey, I.~Biswas, R.~K. Gupta, V.~S. Yadav, A.~Yadav, N.~Yadav,
  H.~Zheng, M.~Henini, A.~Mondal, Cuo–tio2 based self-powered broad band
  photodetector, Nano Materials Science 6~(3) (2024) 345--354.
\newblock \href {https://doi.org/https://doi.org/10.1016/j.nanoms.2023.11.003}
  {\path{doi:https://doi.org/10.1016/j.nanoms.2023.11.003}}.

\bibitem{Mishra-2019}
M.~Mishra, A.~Gundimeda, T.~Garg, A.~Dash, S.~Das, Vandana, G.~Gupta, Zno/gan
  heterojunction based self-powered photodetectors: Influence of interfacial
  states on uv sensing, Applied Surface Science 478 (2019) 1081--1089.
\newblock \href {https://doi.org/https://doi.org/10.1016/j.apsusc.2019.01.192}
  {\path{doi:https://doi.org/10.1016/j.apsusc.2019.01.192}}.

\bibitem{Ghamgosar-2019}
P.~Ghamgosar, F.~Rigoni, M.~G. Kohan, S.~You, E.~A. Morales, R.~Mazzaro,
  V.~Morandi, N.~Almqvist, I.~Concina, A.~Vomiero, Self-powered photodetectors
  based on core–shell zno–co3o4 nanowire heterojunctions, ACS Applied
  Materials \& Interfaces 11~(26) (2019) 23454--23462.
\newblock \href {https://doi.org/10.1021/acsami.9b04838}
  {\path{doi:10.1021/acsami.9b04838}}.

\bibitem{CAO-2021}
F.~Cao, L.~Jin, Y.~Wu, X.~Ji, High-performance, self-powered uv photodetector
  based on au nanoparticles decorated zno/cui heterostructure, Journal of
  Alloys and Compounds 859 (2021) 158383.
\newblock \href {https://doi.org/https://doi.org/10.1016/j.jallcom.2020.158383}
  {\path{doi:https://doi.org/10.1016/j.jallcom.2020.158383}}.

\bibitem{Shen-2015}
Y.~Shen, X.~Yan, Z.~Bai, X.~Zheng, Y.~Sun, Y.~Liu, P.~Lin, X.~Chen, Y.~Zhang, A
  self-powered ultraviolet photodetector based on solution-processed
  p-nio/n-zno nanorod array heterojunction, RSC Adv. 5 (2015) 5976--5981.
\newblock \href {https://doi.org/10.1039/C4RA12535E}
  {\path{doi:10.1039/C4RA12535E}}.

\bibitem{Lin-2020}
H.-P. Lin, P.-Y. Lin, D.-C. Perng,
  \href{https://dx.doi.org/10.1149/1945-7111/ab7e8e}{Fast-response and
  self-powered cu2o/zno nanorods heterojunction uv-visible (570 nm)
  photodetectors}, Journal of The Electrochemical Society 167~(6) (2020)
  067507.
\newblock \href {https://doi.org/10.1149/1945-7111/ab7e8e}
  {\path{doi:10.1149/1945-7111/ab7e8e}}.
\newline\urlprefix\url{https://dx.doi.org/10.1149/1945-7111/ab7e8e}

\bibitem{Benyahia-2021}
K.~Benyahia, F.~Djeffal, H.~Ferhati, A.~Bendjerad, A.~Benhaya, A.~Saidi,
  Self-powered photodetector with improved and broadband multispectral
  photoresponsivity based on zno-zns composite, Journal of Alloys and Compounds
  859 (2021) 158242.
\newblock \href {https://doi.org/https://doi.org/10.1016/j.jallcom.2020.158242}
  {\path{doi:https://doi.org/10.1016/j.jallcom.2020.158242}}.

\end{thebibliography}





\end{document}